# Fast Reconstruction Algorithm for Perturbed Compressive Sensing Based on Total Least-Squares and Proximal Splitting

Reza Arablouei

Commonwealth Scientific and Industrial Research Organisation (CSIRO), Pullenvale, QLD, Australia

**Abstract**: We consider the problem of finding a sparse solution for an underdetermined linear system of equations when the known parameters on both sides of the system are subject to perturbation. This problem is particularly relevant to reconstruction in fully-perturbed compressive-sensing setups where both the projected measurements of an unknown sparse vector and the knowledge of the associated projection matrix are perturbed due to noise, error, mismatch, etc. We propose a new iterative algorithm for tackling this problem. The proposed algorithm utilizes the proximal-gradient method to find a sparse total least-squares solution by minimizing an $\ell_1$-regularized Rayleigh-quotient cost function. We determine the step-size of the algorithm at each iteration using an adaptive rule accompanied by backtracking line search to improve the algorithm's convergence speed and preserve its stability. The proposed algorithm is considerably faster than a popular previously-proposed algorithm, which employs the alternating-direction method and coordinate-descent iterations, as it requires significantly fewer computations to deliver the same accuracy. We demonstrate the effectiveness of the proposed algorithm via simulation results.

## 1. Introduction

The theory of compressive sensing states that an $N$-dimensional vector that has only $K \ll N$ nonzero entries can be recovered from its $M < N$ projections when $M$ is sufficiently set and the projection matrix has certain properties. The projection matrix is generally the product of a sensing matrix and a sparse representation matrix. Conventionally, when reconstructing the unknown sparse vector, the projection matrix is assumed to be perfectly known while the projections may be subject to perturbation stemming from background noise or measurement error [1]-[3]. In many applications, the perfect knowledge of the projection matrix is infeasible and only a perturbed version of it is available for the reconstruction of the original sparse vector. Examples of such applications are grid-based approaches to time-delay/Doppler-shift/direction-of-arrival/position estimation in communications/radar systems or spectrum sensing in cognitive radio networks [4]-[10], mobile electrocardiogram (ECG) monitoring [11], X-ray imaging [12], plant biomass characterization [13], hyperspectral unmixing [14], information security [15], [16], and high-dimensional linear regression [17].

The presence of perturbation in both the projection matrix and the vector of projections has given rise to the so-called perturbed compressive sensing (PCS) paradigm. The reconstruction of the target sparse vector in PCS amounts to solving a fully-perturbed underdetermined system of linear equations (SLE) under a sparsity assumption. The effects of perturbation on the projection matrix as well as the vector of projections have been analysed and relevant theoretical performance bounds have been reported in several works including [18]-[21].

The total least-squares (TLS) method is an effective way of solving a fully-perturbed SLE, albeit disregarding any possible sparseness in the target solution vector. The TLS is a linear fitting technique that accounts for perturbations on both sides of an SLE [22]-[26]. An augmented matrix can be formed by concatenating the perturbed parameter matrix of the left-hand side of an SLE with the perturbed parameter vector of its right-hand side. The optimal TLS solution to the SLE is related to the singular vector of this augmented matrix that corresponds to its smallest singular value [24]. This singular value is the minimum value for the Rayleigh-quotient of the covariance of the augmented matrix [25]-[28]. Besides, it is appreciated that adding an $\ell_1$-norm penalty as a regularization term to a least-squares cost function promotes sparsity in its minimizer [29]. Therefore, one can expect to attain a sparse solution to a fully-perturbed underdetermined SLE by combining the concepts of TLS fitting and $\ell_1$-norm regularization. In [4], two algorithms based on $\ell_1$-regularized TLS estimation are proposed, one of which is near-optimal and the other is suboptimal but with reduced complexity.

The matrix-uncertainty generalized approximate message-passing (MU-GAMP) algorithm, proposed in [30], takes a Bayesian estimation approach to solve the PCS recovery problem. It is based on the generalized approximate message-passing (GAMP) algorithm [31], [32] and exploits the prior knowledge about the probability distribution of the target sparse vector. Another popular approach for reconstruction in PCS is to employ a greedy strategy to enforce sparsity on the TLS (or any other perturbation-compensated) estimate. Some of the greedy PCS reconstruction algorithms combine the TLS estimation with a greedy support-detection method; others modify the classical greedy compressive-sensing reconstruction algorithms, such as the orthogonal matching pursuit algorithm, so that they can account for the perturbation in the projection matrix as well as the perturbation in the vector of projections. The algorithms proposed in [6], [11], [33]-[35] are a few examples. These algorithms along with those based on the $\ell_1$-regularized TLS estimation are among the most computationally efficient reconstruction algorithms for PCS.

In this paper, we propose a new reconstruction algorithm for PCS that finds a sparse total least-squares estimate by minimizing an $\ell_1$-regularized Rayleigh-quotient cost function. To this end, we utilize the proximal-gradient method, which is also known as the forward-backward splitting method, [36]-[38] and determine its step-sizes through an adaptive scheme accompanied by backtracking line search. The proximal-gradient method is suitable for minimizing composite cost functions comprised of two additive terms, one of which is differentiable and the other is convex and admits a proximity operator [39]. It is a two-stage iterative algorithm that addresses each term in the composite cost function separately. At each iteration, it moves the estimate along the opposite direction of the gradient of the differentiable term (forward gradient-descent step), then adjusts the estimate by applying the proximity operator of the other term (backward gradient-descent step) [36]. The proximal-gradient method is a simple practical algorithm that can often be implemented with relatively low complexity.

We assume that the perturbations have Gaussian distribution. The case of perturbations with Poisson distribution has been studied in [12]. Thanks to the computational efficiency of the proximal-gradient method, the computational complexity of the proposed algorithm is of order $\mathcal{O}(NK)$ to $\mathcal{O}(N^2)$ at each iteration, depending on the sparseness of the estimates. Consequently, the proposed algorithm demands significantly fewer computations than its closest contender, the suboptimal algorithm of [4], which has a per-iteration computational complexity order of $\mathcal{O}(NMK)$ to $\mathcal{O}(N^2M)$. Notably, we achieve this improvement in complexity with no sacrifice of estimation accuracy.

We provide simulation examples to examine the performance of the proposed algorithm in comparison with the algorithm of [4]. The simulation results corroborate the efficacy of the proposed algorithm.

## 2. Problem

We consider the problem of finding an estimate of the vector $\mathbf{x}_o \in \mathbb{R}^{N \times 1}$ as the solution of the following underdetermined system of linear equations:

$$\mathbf{A}_o \mathbf{x}_o = \mathbf{b}_o. \tag{1}$$

The matrix $\mathbf{A}_o \in \mathbb{R}^{M \times N}$ has fewer rows than columns, i.e., $M < N$, and the target vector $\mathbf{x}_o$ is sparse with $K < M$ nonzero entries. We do not observe $\mathbf{A}_o$ and $\mathbf{b}_o \in \mathbb{R}^{M \times 1}$ directly. Instead, we have access to their perturbed versions $\mathbf{A}$ and $\mathbf{b}$, respectively, which are given by

$$\mathbf{A} = \mathbf{A}_o - \mathbf{E}_o \text{ and } \mathbf{b} = \mathbf{b}_o - \mathbf{e}_o \tag{2}$$

where $\mathbf{E}_o \in \mathbb{R}^{M \times N}$ and $\mathbf{e}_o \in \mathbb{R}^{M \times 1}$ are unknown perturbations. The source of perturbation can be background noise, measurement error, quantization error, coarse discretization of the parameter space, sampling jitter, mismatch between the true values and the existing knowledge about them, etc. Substituting (2) into (1) gives the following errors-in-variables equation:

$$(\mathbf{A} + \mathbf{E}_o)\mathbf{x}_o = (\mathbf{b} + \mathbf{e}_o).$$

In the context of compressive sensing, $\mathbf{A}_o$ is known as the unperturbed projection or measurement matrix and $\mathbf{b}_o$ is called the vector of unperturbed projections or measurements. The parameters with index $o$ are unobservable (hidden). We are primarily interested in estimating (recovering) the sparse target vector $\mathbf{x}_o$ from the known but perturbed parameters $\mathbf{A}$ and $\mathbf{b}$. As shown in [4], this problem, i.e., reconstruction in PCS, can be cast as an $\ell_1$-regularized total least-squares problem where estimates of $\mathbf{x}_o$, $\mathbf{e}_o$, and $\mathbf{E}_o$ are found by solving

$$\begin{aligned} &\min_{\mathbf{x}, \mathbf{e}, \mathbf{E}} \left( \|\mathbf{e}\|_2^2 + \|\mathbf{E}\|_F^2 + \lambda \|\mathbf{x}\|_1 \right) \\ &\text{subject to } (\mathbf{A} + \mathbf{E})\mathbf{x} = \mathbf{b} + \mathbf{e}. \end{aligned} \tag{3}$$

Here, $\lambda > 0$ is the regularization parameter while $\|\cdot\|_2$, $\|\cdot\|_F$, and $\|\cdot\|_1$ stand for $\ell_2$, Frobenius, and $\ell_1$ norms, respectively. The estimates for $\mathbf{x}_o$, $\mathbf{e}_o$, and $\mathbf{E}_o$ produced by solving (3) are optimal in the maximum *a posteriori* (MAP) sense when the entries of $\mathbf{x}_o$ are independently drawn from a zero-mean common Laplace distribution with parameter $2/\lambda$ and the entries of $\mathbf{e}_o$ and $\mathbf{E}_o$ are independent identically-distributed Gaussian with zero mean and equal variance. According to Lemma 1 of [4], the constrained optimization problem (3) is equivalent to two unconstrained optimization problems, one involving the variables $\mathbf{x}$ and $\mathbf{E}$ as

$$\min_{\mathbf{x},\mathbf{E}}[\|(\mathbf{A}+\mathbf{E})\mathbf{x}-\mathbf{b}\|_2^2 + \|\mathbf{E}\|_F^2 + \lambda\|\mathbf{x}\|_1] \tag{4}$$

and the other involving only $\mathbf{x}$ as

$$\min_{\mathbf{x}}\left(\frac{\|\mathbf{A}\mathbf{x}-\mathbf{b}\|_2^2}{\|\mathbf{x}\|_2^2 + 1} + \lambda\|\mathbf{x}\|_1\right). \tag{5}$$

In [4], two iterative algorithms are proposed for solving the $\ell_1$-regularized total least-squares problem. The first algorithm solves an equivalent form of (5) expressed as

$$\min_{\mathbf{x}} \frac{\|\mathbf{A}\mathbf{x}-\mathbf{b}\|_2^2}{\|\mathbf{x}\|_2^2 + 1} \tag{6}$$
$$\text{subject to } \|\mathbf{x}\|_1 \leq \delta$$

where $\delta$ is an estimate of the $\ell_1$-norm of the optimal solution evaluated through a cross-validation scheme. This algorithm has inner and outer iteration loops that are based on a variant of the branch-and-bound method [40] and the bisection method [41], respectively. Although this algorithm is guaranteed to converge to an arbitrarily small neighbourhood of the global solution of (6), it is computationally expensive as its complexity is not necessarily of polynomial order.

The second algorithm proposed in [4] solves (4) using an alternating-direction approach. It successively alternates between two steps: 1) estimating $\mathbf{x}_o$ given the last estimate of $\mathbf{E}_o$ using coordinate-descent iterations and 2) estimating $\mathbf{E}_o$ given the last estimate of $\mathbf{x}_o$ by solving a straightforward quadratic subproblem. This algorithm, which we will refer to as the alternating-direction coordinate-descent (AD-CD) algorithm, is only guaranteed to converge to a local minimum of the cost function in (4) and not necessarily to its global minimum. However, it is computationally more efficient than the first algorithm. We summarize the AD-CD algorithm in Table I where $\mathbf{a}_{i,n}$ denotes the $i$th column of $\mathbf{A} + \mathbf{E}_n$, which is in fact the estimate of $\mathbf{A}_o$ at iteration $n$. The computational complexity of the AD-CD algorithm is dominated by the calculation of $\mathbf{e}_{i,n}$ for $i = 1, \dots, N$ at each iteration. Therefore, depending on the sparsity level of $\mathbf{x}_n$, the order of the algorithm's per-iteration computational complexity is between $\mathcal{O}(NMK)$ and $\mathcal{O}(N^2M)$.

## 3. Algorithm

In this section, we propose a new algorithm for solving (5). We denote the cost function in (5) by $c(\mathbf{x})$ and split it as

$$c(\mathbf{x}) = f(\mathbf{x}) + r(\mathbf{x})$$

such that

$$f(\mathbf{x}) = \frac{\|\mathbf{A}\mathbf{x}-\mathbf{b}\|_2^2}{\|\mathbf{x}\|_2^2 + 1} \tag{7}$$

and

$$r(\mathbf{x}) = \lambda\|\mathbf{x}\|_1.$$

To minimize the composite cost function $c(\mathbf{x})$, we utilize the proximal-gradient method [36]-[38]. Consequently, the iterations of the proposed algorithm are expressed as

$$\mathbf{z}_n = \mathbf{x}_n - \mu_n \mathbf{g}_n \tag{8}$$

$$\mathbf{x}_{n+1} = \text{prox}_{\mu_n r}\{\mathbf{z}_n\} \tag{9}$$

where $\mathbf{x}_n$ is the estimate of $\mathbf{x}_o$ at iteration $n$, $\mu_n$ is the step-size at iteration $n$, $\mathbf{g}_n$ is the gradient of $f(\mathbf{x})$ at $\mathbf{x}_n$, $\mathbf{z}_n$ is the intermediate estimate, and $\text{prox}_{\mu_n r}\{\mathbf{z}_n\}$ denotes the proximity operator of the function $\mu_n r(\mathbf{x}) = \mu_n \lambda\|\mathbf{x}\|_1$ at point $\mathbf{z}_n$, defined as

$$\text{prox}_{\mu_n r}\{\mathbf{z}_n\} = \arg\min_{\mathbf{u}}\left[\mu_n r(\mathbf{u}) + \frac{1}{2}\|\mathbf{z}_n - \mathbf{u}\|_2^2\right].$$

This operator is separable; hence, it applies to the entries of $\mathbf{z}_n$ independently. Let $z_{i,n}$ denote the $i$th entry of $\mathbf{z}_n$. Thus, we get

$$\text{prox}_{\mu_n r}\{\mathbf{z}_n\} = \begin{bmatrix} \text{prox}_{\mu_n r}\{z_{1,n}\} \\ \vdots \\ \text{prox}_{\mu_n r}\{z_{N,n}\} \end{bmatrix}.$$

The proximity operator of the absolute-value function is the shrinkage operator. Therefore, we have

$$\text{prox}_{\mu_n r}\{z_{i,n}\} = \begin{cases} z_{i,n} - \mu_n \lambda & z_{i,n} > \mu_n \lambda \\ 0 & |z_{i,n}| \leq \mu_n \lambda \\ z_{i,n} + \mu_n \lambda & z_{i,n} < -\mu_n \lambda. \end{cases}$$

The gradient of $f(\mathbf{x})$ is calculated as

$$\mathbf{g}(\mathbf{x}) = \frac{2}{(\|\mathbf{x}\|_2^2 + 1)^2} [(\|\mathbf{x}\|_2^2 + 1)\mathbf{A}^\top(\mathbf{A}\mathbf{x} - \mathbf{b}) - \|\mathbf{A}\mathbf{x} - \mathbf{b}\|_2^2 \mathbf{x}]$$

and its value at $\mathbf{x}_n$ is

$$\mathbf{g}_n = \frac{2}{(\|\mathbf{x}_n\|_2^2 + 1)^2} [(\|\mathbf{x}_n\|_2^2 + 1)\mathbf{A}^\top(\mathbf{A}\mathbf{x}_n - \mathbf{b}) - \|\mathbf{A}\mathbf{x}_n - \mathbf{b}\|_2^2 \mathbf{x}_n].$$

For the step-size $\mu_n$, we use the adaptive rule presented in [39], which is a hybrid between the spectral method of [42] and the adaptive step-size scheme of [43]. According to this rule, two parameters, called the "steepest descent" step-size and the "minimum residual" step-size, are calculated as

$$\mu_{s,n} = \frac{\|\mathbf{x}_n - \mathbf{x}_{n-1}\|_2^2}{(\mathbf{x}_n - \mathbf{x}_{n-1})^\top (\mathbf{g}_n - \mathbf{g}_{n-1})}$$

and

$$\mu_{m,n} = \frac{(\mathbf{x}_n - \mathbf{x}_{n-1})^\top (\mathbf{g}_n - \mathbf{g}_{n-1})}{\|\mathbf{g}_n - \mathbf{g}_{n-1}\|_2^2},$$

respectively. Then, the step-size of iteration $n$ is computed as

$$\mu_n = \begin{cases} \mu_{m,n} & \frac{\mu_{m,n}}{\mu_{s,n}} > 0.5 \\ \mu_{s,n} - \frac{\mu_{m,n}}{2} & \text{otherwise.} \end{cases}$$

If the value of $\mu_n$ becomes negative or zero, we replace it with the step-size of the previous iteration, $\mu_{n-1}$. In order to ensure stability of the algorithm, we use backtracking line search to curb the value of the step-size into its stable operating region [44]. For this purpose, we check the following line-search condition after the proximal-gradient steps (8) and (9):

$$f(\mathbf{x}_{n+1}) < f(\mathbf{x}_n) + (\mathbf{x}_{n+1} - \mathbf{x}_n)^\top \mathbf{g}_n + \frac{1}{2\mu_n} \|\mathbf{x}_{n+1} - \mathbf{x}_n\|_2^2. \qquad (10)$$

If it is violated, we reduce $\mu_n$ by half and repeat the proximal-gradient steps. Choosing any value for $\mu_n$ that is smaller than the reciprocal of the Lipschitz constant of $\mathbf{g}(\mathbf{x})$ guarantees the satisfaction of the condition (10).

We summarize the proposed algorithm in Table II. Note that the matrix-matrix and matrix-vector products $\mathbf{A}^\top\mathbf{A}$ and $\mathbf{A}^\top\mathbf{b}$ are not performed at every iteration but they are pre-computed at the initialization phase. The computational complexity of the proposed algorithm is dominated by the steps involving the calculation of the matrix-vector multiplications $\mathbf{A}^\top\mathbf{A}\mathbf{x}_n$ and $\mathbf{A}\mathbf{x}_{n+1}$. Therefore, depending on the level of sparseness of the estimates $\mathbf{x}_n$ and $\mathbf{x}_{n+1}$, the order of the per-iteration computational complexity of the proposed algorithm is between $\mathcal{O}(NK)$ and $\mathcal{O}(N^2)$. Thus, each iteration of the proposed algorithm requires nearly an order of magnitude fewer computations compared with the AD-CD algorithm.

The optimization problem (5) is non-convex since $f(\mathbf{x})$, defined by (7), is not convex on its entire domain $\mathbb{R}^{N \times 1}$. Hence, the global convergence of the proposed algorithm from an arbitrary initial estimate cannot be guaranteed. However, following similar arguments as in [26], it is possible to verify that the proposed algorithm is at least locally convergent. In other words, the proposed algorithm converges when the initial estimate is sufficiently close to a solution of (5). This would of course be a conservative theoretical conclusion. In practice, we have observed that, the proposed algorithm converges with a wide range of initial estimates. Moreover, to ensure a good initialization, one might first solve the associated $\ell_1$-regularized least-squares problem, which ignores the perturbation in the projection matrix, and use the solution as the initial estimate for the proposed algorithm. We will present the detailed performance analysis of the proposed algorithm in another publication.

## 5. Simulations

In this section, we examine the performance of the proposed algorithm in comparison with the AD-CD algorithm of [4] in two example scenarios. In both scenarios, the non-zero entries of $\mathbf{x}_o$ have arbitrary values and are normalized to yield a unit $\ell_2$-norm for $\mathbf{x}_o$. The entries of $\mathbf{E}_o$ and $\mathbf{e}_o$ are independently drawn from a Gaussian distribution with zero mean and variance $\xi/M$. We call $\xi$ the perturbation-variance parameter. In the first scenario, we set $N = 40$, $M = 20$, $K = 5$, and draw the entries of $\mathbf{A}_o$ independently from a Gaussian distribution with zero mean and variance $1/M$. In the second scenario, we set $N = 200$, $M = 80$, $K = 20$, and draw the entries of $\mathbf{A}_o$ independently from a Rademacher distribution with zero mean and variance $1/M$. Hence, the entries of $\mathbf{A}_o$ take the values of $-1/\sqrt{M}$ or $+1/\sqrt{M}$ with equal chance. Note that, in both scenarios, the columns of $\mathbf{A}_o$ have unit expected $\ell_2$-norm. We run the algorithms for a pre-determined number of iterations depending on the value of the regularization parameter $\lambda$. The smaller the value of $\lambda$ is, the more iterations are required for convergence. In our experiments, we use values for $\lambda$ that range from $5 \times 10^{-4}$ to 1. Accordingly, we vary the number of iterations logarithmically between 2800 and 40 in the first scenario and between 3500 and 50 in the second scenario, as shown in Fig. 1. Our experiments show that both considered algorithms converge adequately with these iteration numbers. We obtain the simulation results by averaging over 100 independent trials.

Fig. 2 shows how the squared-$\ell_2$-norm of the reconstruction error, i.e., $\|\mathbf{x}_n - \mathbf{x}_o\|_2^2$, evolves in time when $\lambda = 0.02$ and $\xi = 0.01$. Fig. 3 displays the converged squared-$\ell_2$-norm of the reconstruction error, i.e., $\|\mathbf{x}_\infty - \mathbf{x}_o\|_2^2$, as a function of $\lambda$ when $\xi = 0.01$. Here, $\mathbf{x}_\infty$ denotes the converged value of $\mathbf{x}_n$, i.e., its value at the last iteration. Fig. 4 depicts the rates of false-negative as well as false-positive support-detection errors versus $\lambda$ when $\xi = 0.01$. The false-negative support-detection error is the number of non-zero entries of $\mathbf{x}_o$ that are incorrectly estimated as zero in $\mathbf{x}_\infty$ and the false-positive support-detection error is the number of non-zero entries of $\mathbf{x}_\infty$ whose corresponding entries in $\mathbf{x}_o$ are zero. Fig. 5 plots the converged squared-$\ell_2$-norm of the reconstruction error against the perturbation-variance parameter $\xi$ when $\lambda = 0.02$. Figs. 2-5 include the simulation results for both the proposed and AD-CD algorithms in both considered scenarios. The legend $N = 40$ refers to the first scenario and the legend $N = 200$ to the second scenario.

Fig 6 exhibits a comparison of the proposed algorithm and the AD-CD algorithm in terms of the actual running time. In this figure, the horizontal axes are $\lambda$ and the vertical axes are the ratio of the average time elapsed during each iteration of the AD-CD algorithm to that of the proposed algorithm. We obtained these results using MATLAB on a Mobile Workstation with a 2.9GHz Core-i7 CPU and 24GB of DDR3 RAM. As seen in Fig. 6, the proposed algorithm is about 34 times faster than the AD-CD algorithm in the first scenario and about 235 times faster in the second scenario.

The above experiments demonstrate that the estimation accuracy and convergence speed of the proposed algorithm is very similar to that of the AD-CD algorithm. This is evident from the similarity of the results provided in Figs. 2-5 for the AD-CD and the proposed algorithms in both considered scenarios. However, as attested to by the results presented in Fig. 6, the proposed algorithm is significantly faster than the AD-CD algorithm in terms of the actual runtime as it requires considerably fewer computations per iteration.

## 6. Conclusion

We proposed a new algorithm for solving an underdetermined system of linear equations when the solution vector is expected to be sparse and the known parameters on both sides of the system are subject to perturbation. This is particularly relevant to the notion of perturbed compressive sensing where one is interested in reconstructing a sparse vector when only perturbed versions of the projection matrix and the vector of projections are available. This problem can be recast as a sparse total least-squares estimation problem with an $\ell_1$-regularized Rayleigh quotient cost function. We proposed to minimize this cost function using the proximal-gradient method enhanced by an adaptive scheme for choosing the step-size at each iteration. The resultant algorithm requires an order of magnitude fewer computations compared with a previously-proposed algorithm, which uses an alternating-direction method together with coordinate-descent iterations, while the two algorithms have very similar convergence rate and estimation accuracy. We verified the merits of the proposed algorithm through numerical examples.

Table 1. The alternating-direction coordinate-descent (AD-CD) algorithm of [4].

---
initialization
$\quad \mathbf{x}_0 = \mathbf{0}_N$
$\quad \mathbf{E}_0 = \mathbf{0}_{M \times N}$
for $n = 0, 1, \ldots$
$\quad$ for $i = 1, \ldots, N$
$$\mathbf{e}_{i,n} = \mathbf{b} - \sum_{j=1}^{i-1} \mathbf{a}_{j,n} x_{j,n+1} - \sum_{j=i+1}^{N} \mathbf{a}_{j,n} x_{j,n}$$
$$x_{i,n+1} = \frac{1}{\|\mathbf{a}_{i,n}\|_2^2} \begin{cases} \mathbf{e}_{i,n}^\top \mathbf{a}_{i,n} - \lambda/2 & \mathbf{e}_{i,n}^\top \mathbf{a}_{i,n} > \lambda/2 \\ 0 & |\mathbf{e}_{i,n}^\top \mathbf{a}_{i,n}| < \lambda/2 \\ \mathbf{e}_{i,n}^\top \mathbf{a}_{i,n} + \lambda/2 & \mathbf{e}_{i,n}^\top \mathbf{a}_{i,n} < -\lambda/2 \end{cases}$$
$$\mathbf{E}_{n+1} = (\|\mathbf{x}_{n+1}\|_2^2 + 1)^{-1} (\mathbf{b} - \mathbf{A}\mathbf{x}_{n+1}) \mathbf{x}_{n+1}^\top$$

---

Table 2. The proposed algorithm.

---
initialization
$\quad \mathbf{x}_0 = \mathbf{0}_N$
$\quad \mathbf{g}_0 = -2\mathbf{A}^\top \mathbf{b}$
$\quad \mu_0 = 0.2$
$\quad \mathbf{x}_1 = \text{prox}_{\mu_0 r}\{\mathbf{x}_0 - \mu_0 \mathbf{g}_0\}$
$\quad y_1 = \dfrac{1}{\|\mathbf{x}_1\|_2^2 + 1}$
$\quad f_1 = y_1 \|\mathbf{A}\mathbf{x}_1 - \mathbf{b}\|_2^2$
for $n = 1, 2, \ldots$
$\quad \mathbf{g}_n = 2y_n (\mathbf{A}^\top \mathbf{A} \mathbf{x}_n - \mathbf{A}^\top \mathbf{b} - f_n \mathbf{x}_n)$
$\quad$ if $(\mathbf{x}_n - \mathbf{x}_{n-1})^\top (\mathbf{g}_n - \mathbf{g}_{n-1}) = 0$ or $\|\mathbf{g}_n - \mathbf{g}_{n-1}\|_2^2 = 0$
$\quad\quad \mu_n = \mu_{n-1}$
$\quad$ else
$$\mu_{s,n} = \frac{\|\mathbf{x}_n - \mathbf{x}_{n-1}\|_2^2}{(\mathbf{x}_n - \mathbf{x}_{n-1})^\top (\mathbf{g}_n - \mathbf{g}_{n-1})}$$
$$\mu_{m,n} = \frac{(\mathbf{x}_n - \mathbf{x}_{n-1})^\top (\mathbf{g}_n - \mathbf{g}_{n-1})}{\|\mathbf{g}_n - \mathbf{g}_{n-1}\|_2^2}$$
$\quad\quad$ if $\mu_{m,n}/\mu_{s,n} > 0.5$
$\quad\quad\quad \mu_n = \mu_{m,n}$
$\quad\quad$ else
$\quad\quad\quad \mu_n = \mu_{s,n} - \mu_{m,n}/2$
$\quad$ if $\mu_n \leq 0$
$\quad\quad \mu_n = \mu_{n-1}$
$\quad$ repeat
$\quad\quad \mathbf{x}_{n+1} = \text{prox}_{\mu_n r}\{\mathbf{x}_n - \mu_n \mathbf{g}_n\}$
$\quad\quad y_{n+1} = \dfrac{1}{\|\mathbf{x}_{n+1}\|_2^2 + 1}$
$\quad\quad f_{n+1} = y_{n+1} \|\mathbf{A}\mathbf{x}_{n+1} - \mathbf{b}\|_2^2$
$\quad\quad$ if $f_{n+1} < f_n + (\mathbf{x}_{n+1} - \mathbf{x}_n)^\top \mathbf{g}_n + \frac{1}{2\mu_n} \|\mathbf{x}_{n+1} - \mathbf{x}_n\|_2^2$
$\quad\quad\quad$ break
$\quad\quad$ else
$\quad\quad\quad \mu_n = \mu_n / 2$

---

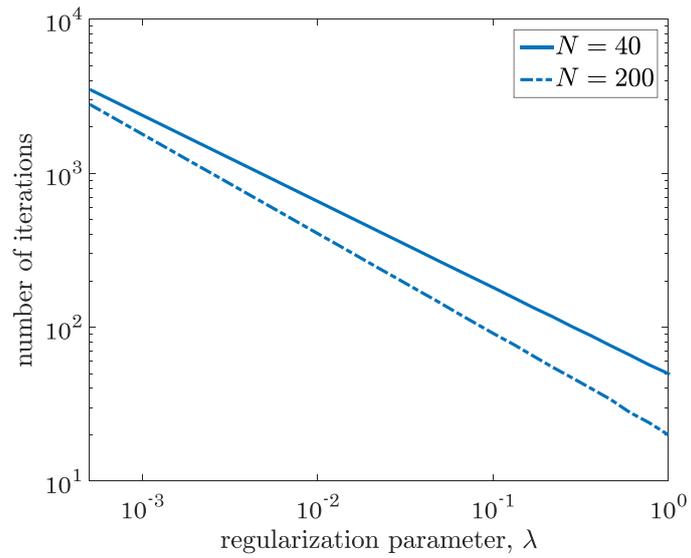

Fig. 1. The number of iterations for which both considered algorithms run depending on the value of the regularization parameter $\lambda$. The curve with the legend $N = 40$ corresponds to the first scenario ($N = 40$, $M = 20$, $K = 5$, Gaussian $\mathbf{A}_o$) and the curve with the legend $N = 200$ corresponds to the second scenario ($N = 200$, $M = 80$, $K = 20$, Rademacher $\mathbf{A}_o$).

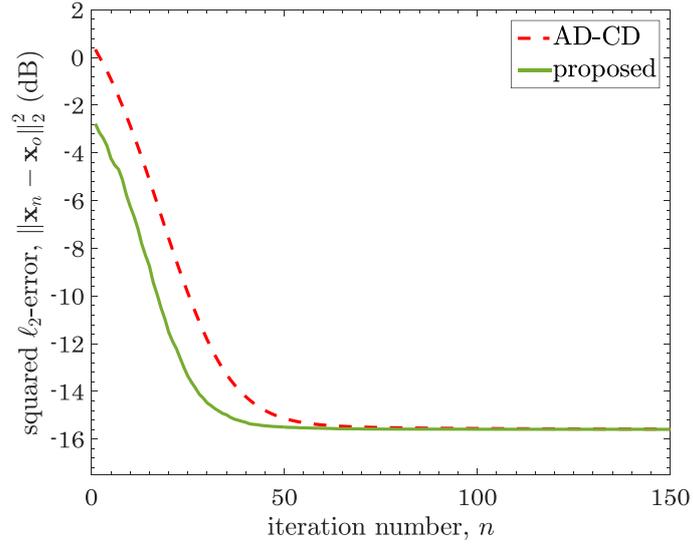

(a)

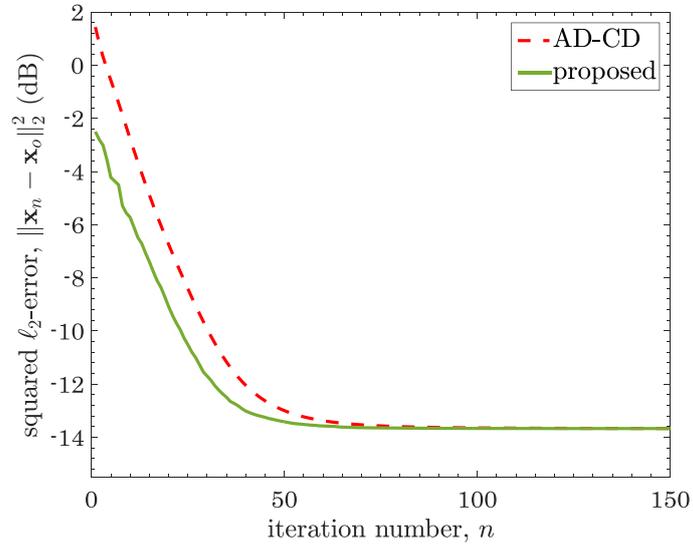

(b)

Fig. 2. The squared-$\ell_2$-norm of the reconstruction error at different iterations when $\lambda = 0.02$ and $\xi = 0.01$ in the first (a) and the second (b) scenarios. The AD-CD and proposed algorithms converge to very similar values in both considered scenarios.

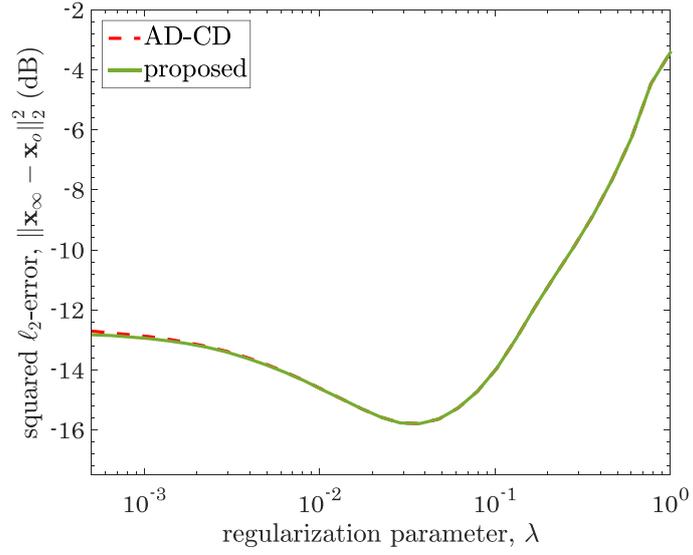

(a)

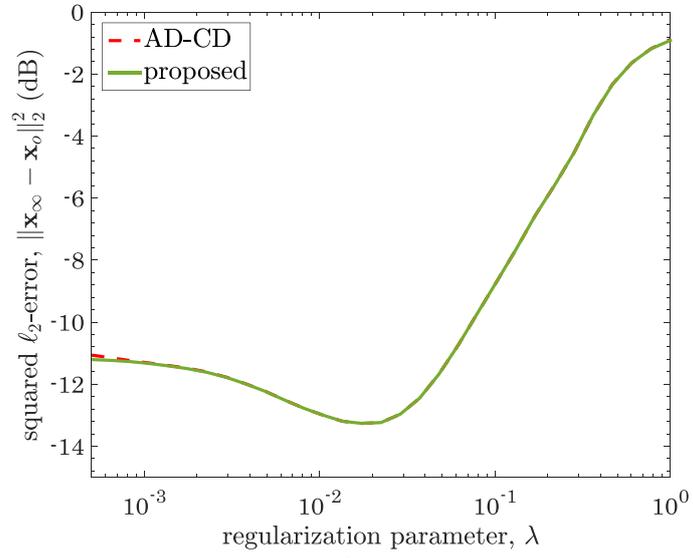

(b)

Fig. 3. The converged squared-$\ell_2$-norm of the reconstruction error for different values of $\lambda$ when $\xi = 0.01$ in the first (a) and the second (b) scenarios. The results for the AD-CD and proposed algorithms are very similar for most values of $\lambda$ in both considered scenarios.

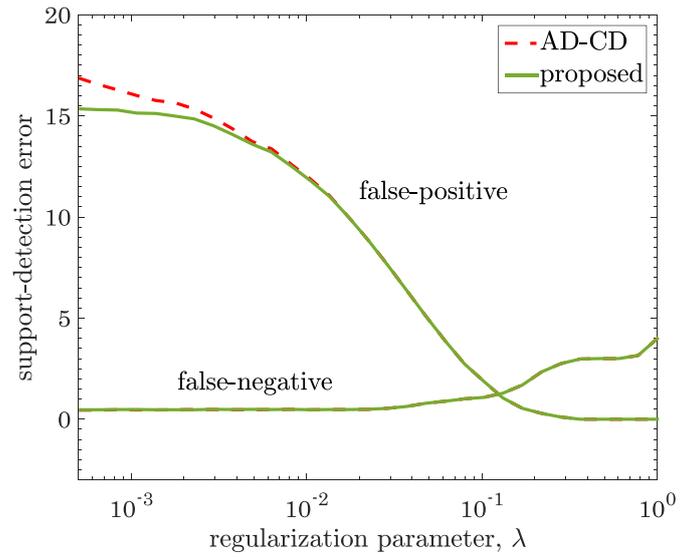

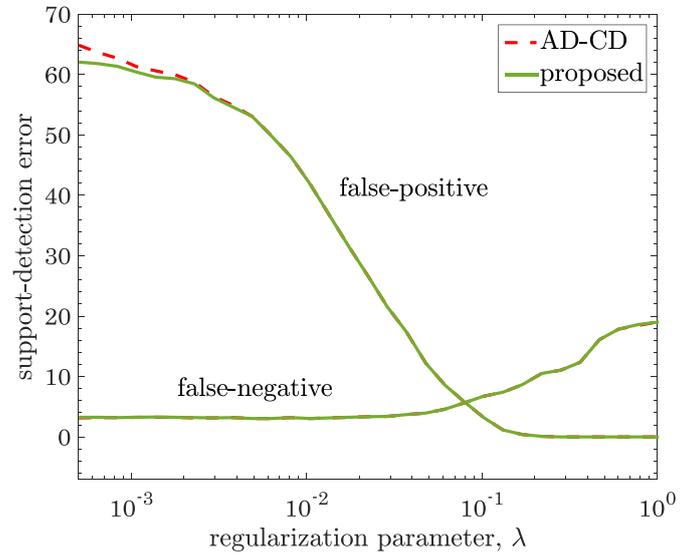

Fig. 4. The rates of false-negative and false-positive support-detection errors for different values of $\lambda$ when $\xi = 0.01$ in the first (a) and the second (b) scenarios. In both scenarios, the false-negative errors of the AD-CD and proposed algorithms are almost identical for any value of $\lambda$ in the considered range and the false-positive errors are almost identical when $\lambda > 0.02$.

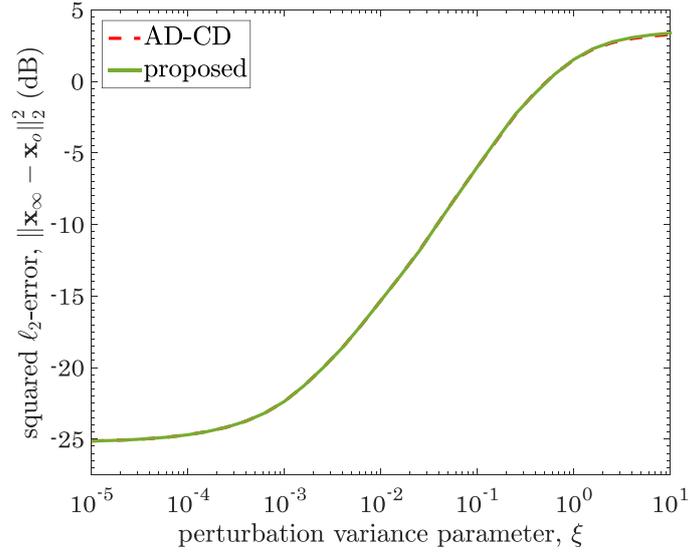

(a)

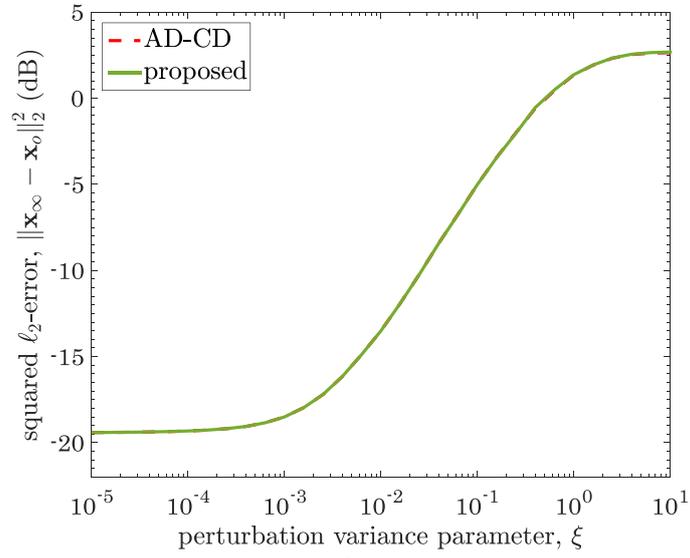

(b)

Fig. 5. The converged squared-$\ell_2$-norm of the reconstruction error for different values of $\xi$ when $\lambda = 0.02$ in the first (a) and the second (b) scenarios. The results for the AD-CD and proposed algorithms are very similar for all considered values of $\xi$ in both considered scenarios.

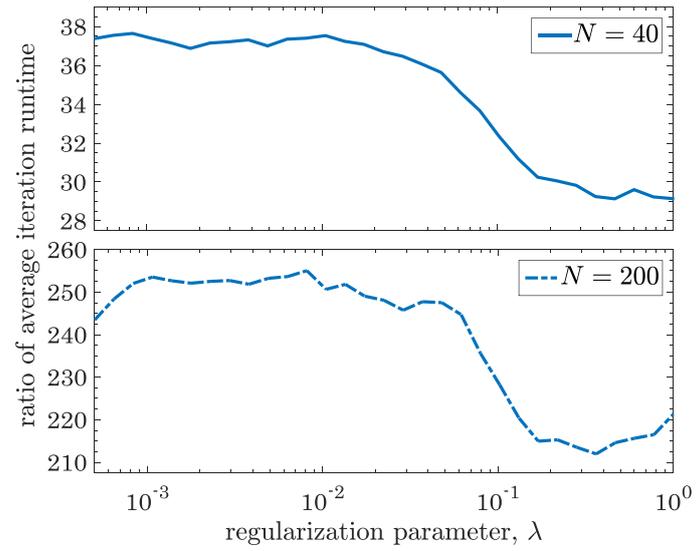

Fig. 6. The average running time of each iteration of the AD-CD algorithm divided by the average running time of each iteration of the proposed algorithm for different values of λ in both considered scenarios.